\documentclass{aip-cp}

\usepackage[numbers]{natbib}
\usepackage{rotating}
\usepackage{graphicx}
\usepackage{mathrsfs}
\usepackage{subfigure}

\begin{document}

\title{Higher Order Cumulants in Colorless Partonic Plasma}

\author[aff2,aff4]{S. CHERIF}
\author[aff1,aff3,aff4]{M.A.A. AHMED}
\author[aff1,aff4]{M. LADREM\corref{cor1}}

\affil[aff1]{Department of Physics, College of Science,Taibah University Al-Madinah Al-Mounawwarah KSA}
\affil[aff2]{Sciences and Technologies Department,University of Ghardaia,Ghardaia, Algiers}
\affil[aff3]{Department of Physics,Taiz University in Turba,Taiz ,Yemen}
\affil[aff4]{\textbf{L}aboratoire de \textbf{P}hysique et de \textbf{M}ath\'ematiques \textbf{A}ppliqu\'ees (LPMA),ENS-Kouba (Bachir El-Ibrahimi),Algiers,Algeria.}
\corresp[cor1]{Corresponding author: mladrem@yahoo.fr}

\maketitle

\begin{abstract}
Any physical system considered to study the QCD deconfinement phase transition certainly has a finite volume, so the finite size effects are inevitably present. This renders the location of the phase transition and the determination of its order as an extremely difficult task, even in the simplest known cases. In order to identify and locate the colorless QCD deconfinement transition point in finite volume $T_{0}(V)$, a new approach based on the finite-size cumulant expansion of the order parameter and the $\mathscr{L}_{m,n}$-Method is used.We have shown that both cumulants of higher order and their ratios, associated to the thermodynamical fluctuations of the order parameter, in QCD deconfinement phase transition behave in a particular enough way revealing pronounced oscillations in the transition region. The sign structure and the oscillatory behavior of these in the vicinity of the deconfinement phase transition point might be a sensitive probe and may allow one to elucidate their relation to the QCD phase transition point. In the context of our model, we have shown that the finite volume transition point is always associated to the appearance of a particular point in whole higher order cumulants under consideration.
\end{abstract}
\section{Introduction}
Despite the creation of the new state and its identification as being the QCD partonic plasma in Ultra-relativistic heavy ion collisions experiments at RHIC and LHC \cite{Qin2014} , the physics of the deconfinement phase transition continues to attract a lot of research interest.
When we insert the colorless condition in the MIT bag model \cite{MITBM} , we obtain the colorless-MIT bag model using a mixed phase system evolving in a finite total volume $V$ \cite{Ladrem2005,Herbadji2007,LZH2011,Mhamed2014,Ladrem2015} . The fraction of volume occupied by the hadronic gas (HG) phase is given by $V_{HG}=\mathbf{h}V,$ and then the remaining volume  $V_{CPP}=(1-\mathbf{h)}V$ contains the colorless partonic plasma phase (CPP). In the case of a non-interacting phases, the total partition function of the system can be written in a simple form,
\begin{equation}
\mathscr{Z}_{TOT}\left(\mathbf{h}\right) =\mathscr{Z}_{HG}(\mathbf{h})\mathscr{Z}_{Vac}(\mathbf{h})\mathscr{Z}_{PP}(\mathbf{h})\mathscr{Z}_{CC}\left( \mathbf{h}\right)
\end{equation}
where $\mathscr{Z}_{CC}\left( \mathbf{h}\right)$ is the colorless part of the partition function and $\mathscr{Z}_{PP}\left( \mathbf{h}\right)$ is the partition function of the partonic plasma. $\mathscr{Z}_{Vac}(\mathbf{h},V,T)$ accounts for the confinement of quarks and gluons by the real vacuum pressure exerted on the perturbative vacuum of the bag model. The hadronic  partition function $\mathscr{Z}_{HG}(\mathbf{h},V,T)$ is just calculated for a pionic gas. The final expressions of these partition functions can be found in \cite{Ladrem2005,Herbadji2007,LZH2011,Mhamed2014,Ladrem2015} .
Any physical system considered to study the QCD deconfinement phase transition certainly has a finite volume, so the finite size effects are inevitably present. This renders the location of the phase transition and the determination of its order as an extremely difficult task, even in the simplest known cases. In order to identify and locate the colorless QCD deconfinement transition point in finite volume $T_{0}(V)$, a new approach based on the finite-size cumulant expansion of the order parameter and the $\mathscr{L}_{m,n}$-Method is used (see details in  \cite{Mhamed2014,Ladrem2015} ). It has been put into evidence that all cumulants and their ratios showed deviations from their asymptotic values, which increase with the cumulant order. This behavior is essential to discriminate the phase transition by measuring the fluctuations. The sign structure and the oscillatory behavior of these in the vicinity of the finite volume transition point might be a sensitive probe and may allow one to elucidate their relation to the QCD phase transition point. In view of this, Higher Order Cumulants (H.O.C) are often used in showing some important physical properties  as well as to look for the position of the finite volume transition point.
\section{Moments, Centered Moments and Cumulants}
The definition of the hadronic probability density function (hpdf) $p(\mathbf{h})$ in our model is given by $p(\mathbf{h})\int\limits_{0}^{1}\mathscr{Z}_{TOT}(\mathbf{h})d%
\mathbf{h}=\mathscr{Z}_{TOT}(\mathbf{h})$. Since our hpdf is directly related to the partition function of the system, it is believed that the whole information concerning the deconfinement phase transition is self-contained in this hpdf. This hpdf should certainly have different behavior in both sides of the phase transition and then we should be able to locate the transition point just by analyzing some of its basic properties. The $n^{th}$ moment of $p(\mathbf{h})$ of the order parameter $\mathbf{h}$ is the mean value of $\mathbf{h}^{n}$. From its definition and after some algebra, we get the general expression of it as a function of only a certain double integral coefficient: $\mathscr{L}_{mn}(1-\mathbf{h},T,V)$ (see details in \cite{Mhamed2014,Ladrem2015} ),
\begin{equation}
\left\langle \mathbf{h}^{n}\right\rangle (T,V) =\int\limits_{0
}^{1}\mathbf{h}^{n}p(\mathbf{h})d\mathbf{h}=\frac{n!\mathscr{L}_{0,n+1}\left(
1,T,V\right) -\sum\limits_{k=0}^{n} {n \choose k} k!\mathscr{L}_{0,k+1}\left( 0,T,V\right) }{\mathscr{L}_{0,1}\left(
1,T,V\right) -\mathscr{L}_{0,1}\left( 0,T,V\right) }. \label{meanhnLmn}  \label{moment}
\end{equation}
Relatively to the mean value $\left\langle \mathbf{h}\right\rangle$, we can define the centered moments $\mathcal{M}_{n}$ as follows,
\begin{equation}
\mathcal{M}_{n}(T,V)=\int\limits_{0 }^{+1 }\left( \mathbf{h}-\left\langle \mathbf{h}\right\rangle%
\right) ^{n}p(\mathbf{h})d\mathbf{h}=\sum_{k=0}^{n}(-1)^{k}{n \choose k}%
\left\langle \mathbf{h}\right\rangle^{k} \left\langle \mathbf{h}^{n-k}\right\rangle,  \label{CMvsM}
\end{equation}
where ${n \choose k}=\frac{n!}{k!(n-k)!}$ are the standard binomial coefficients.
The Cumulants $\mathbf{C}_{n}$ can be computed from the Maclaurin development of the characteristic function of our hpdf \cite{Mhamed2014,Ladrem2015} .
We can write these cumulants in terms of centered moments, which can be combined into a single recursive relationship : $\mathbf{C}_{n}=\mathcal{M}_{n}-\sum\limits_{m=1}^{n-1}{n-1 \choose m-1}\mathbf{C}_{m}\mathcal{M}_{n-m}$. Generically, the structures of all cumulants are related to each other and the behavior including the magnitudes can be deduced from the preceding.
Afterwards, one can express the different cumulants $\mathbf{C}_{n}(T,V)$ in terms of these $\mathscr{L}_{mn}(1-\mathbf{h},T,V)$ using (\ref{meanhnLmn}).
Keeping in mind that these double integrals $\mathscr{L}_{mn}(1-\mathbf{h},T,V)$ are state functions depending on the temperature $T$,
volume $V$ and on the state variable $\mathbf{h}$. General expressions for the connection between cumulants and moments may be found in \cite{Risken1989}.
The cumulants are considered as important quantities in physics but cumulant ratios are more important.
\section{Higher Order Cumulants: Hexosis, Heptosis and Octosis}
More recently \cite{Mhamed2014,Ladrem2015}, a new theoretical formalism  that provides the definitions of cumulant ratios in a clear, unified and consistent way has been developed which also makes predictions for the behavior of H.O.C and their ratios.
We start from the generalized connected cumulant ratios between the cumulants as defined in \cite{Mhamed2014,Ladrem2015}, to rewrite the most useful one which represents the p\textnormal{\textit{th}}-Order Under-Normalized Cumulant Ratios: $\mathcal{K}_{\leq p}^{\{(i,\alpha_{i}\neq0)\}}={\mathbf{C}_{p}}{\prod\limits_{i=1}^{p} \mathbf{C}_{i}^{-\alpha_{i}}}$, with the condition $\sum\limits_{i=1}^{p} \alpha_{i}\times (i)= p.$
The sixth order under-normalized cumulant ratio was coined Hexosis. It was defined in one of the following ways :
\begin{equation}
\mathcal{H}_{1}=\frac{\mathbf{C}_{6}}{\left( \mathbf{C}_{2}\right)
^{3}}, \quad
\mathcal{H}_{2}=\frac{\mathbf{C}_{6}}{\left(\mathbf{C}_{3}\right) ^{2}},\quad    \mathcal{H}_{3}=\frac{\mathbf{C}_{6}}{\mathbf{C}_{4}\mathbf{C}_{2}}.
\label{H123}
\end{equation}
With the same spirit and by analogy to previous appellations, the seventh order under-normalized cumulant ratio was termed Heptosis  $\eta$.
One of the possible definition of heptosis is given by,
\begin{equation}
\eta_1=
\frac{\mathbf{C}_{7}}{\mathbf{C}_{2}^{7/2}},\quad
\eta_2=
\frac{\mathbf{C}_{7}}{\mathbf{C}_{2}\mathbf{C}_{5}},\quad
\eta_3=
\frac{\mathbf{C}_{7}}{\mathbf{C}_{2}^{2}\mathbf{C}_{3}},\quad
\eta_4=
\frac{\mathbf{C}_{7}}{\mathbf{C}_{3}\mathbf{C}_{4}}.\quad
\label{heptosis}
\end{equation}
Finally, the eighth order under-normalized cumulant ratio, which was also called Octosis, is given by one of the following forms
\begin{equation}
\omega_1=\frac{\mathbf{C}_{8}}{\mathbf{C}_{2}^{4}},\quad
\omega_2=\frac{\mathbf{C}_{8}}{\mathbf{C}_{2}^{2}\mathbf{C}_{4}},\quad
\omega_3=\frac{\mathbf{C}_{8}}{\mathbf{C}_{2}\mathbf{C}_{3}^{2}},\quad
\omega_4=\frac{\mathbf{C}_{8}}{\mathbf{C}_{2}\mathbf{C}_{6}},\quad
\omega_5=\frac{\mathbf{C}_{8}}{\mathbf{C}_{3}\mathbf{C}_{5}},\quad
\omega_6=\frac{\mathbf{C}_{8}}{\mathbf{C}_{4}^{2}}. \quad
\label{Octosis}
\end{equation}

\begin{figure}[h]
  \includegraphics[width=250pt]{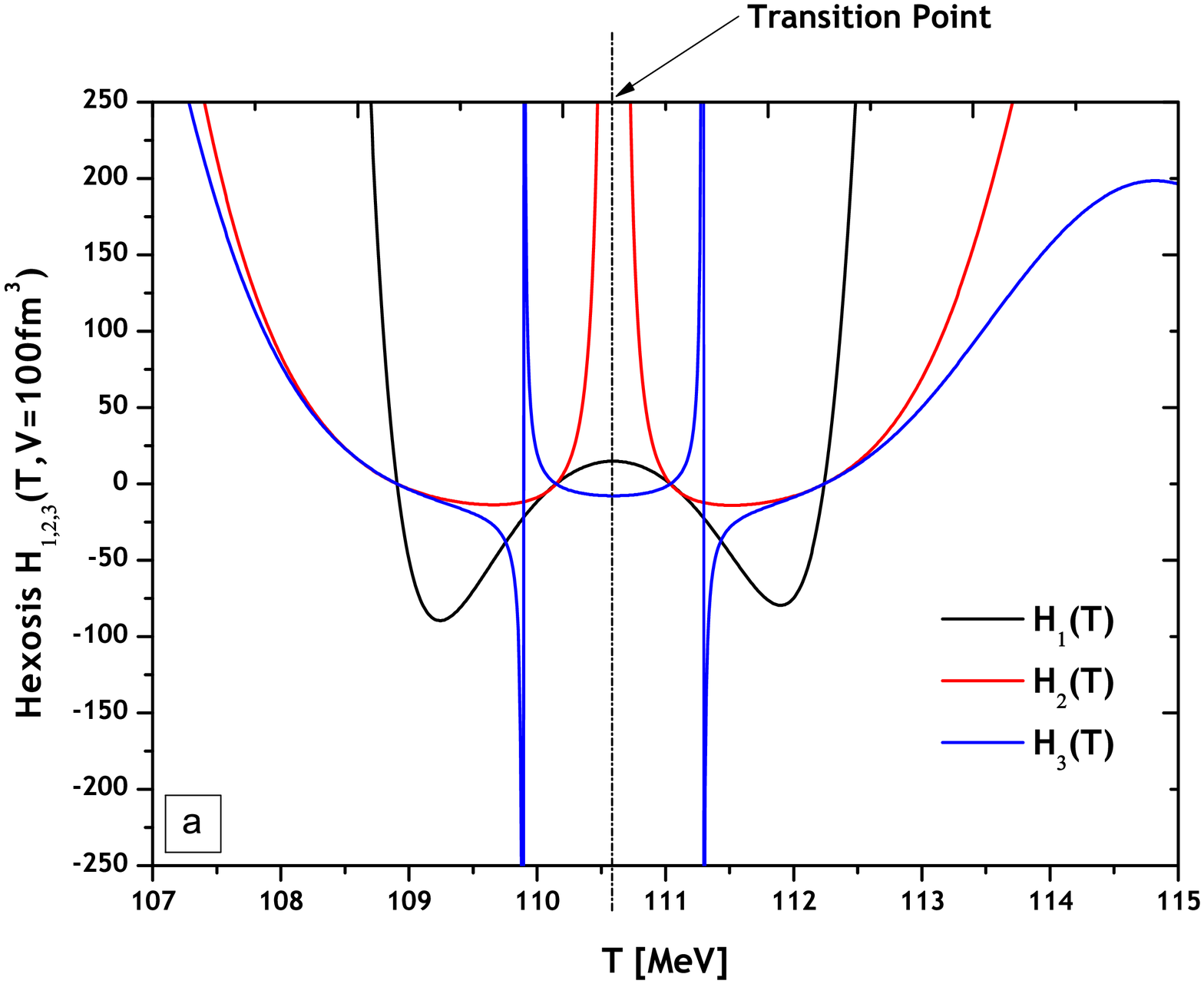}
  \includegraphics[width=250pt]{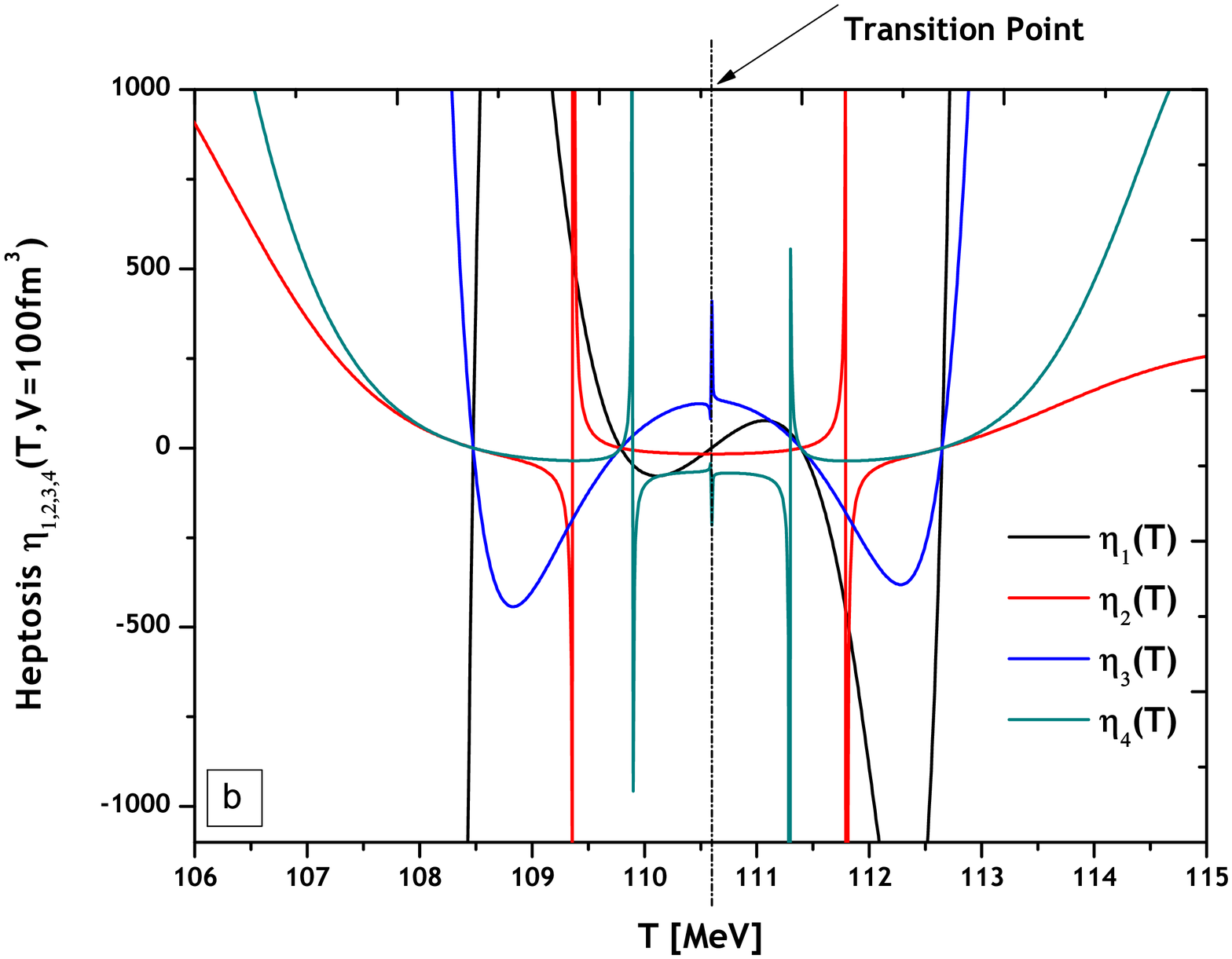}
  \caption{(a) Hexosis vs Temperature at $V=100fm^{3}$ - (b) Heptosis vs Temperature at $V=100fm^{3}$}
\end{figure}

\begin{figure}[h]
  \includegraphics[width=250pt]{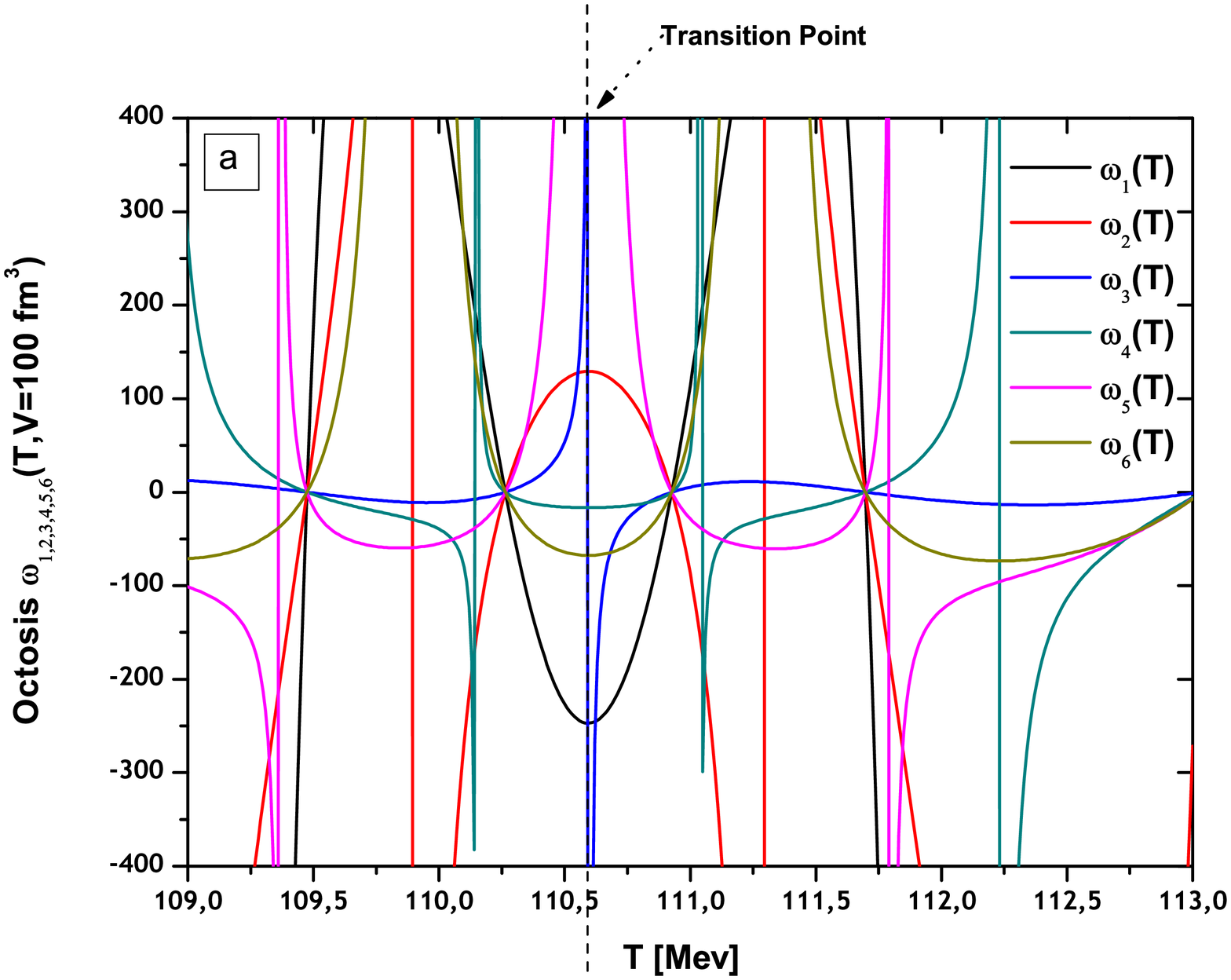}
  \includegraphics[width=250pt]{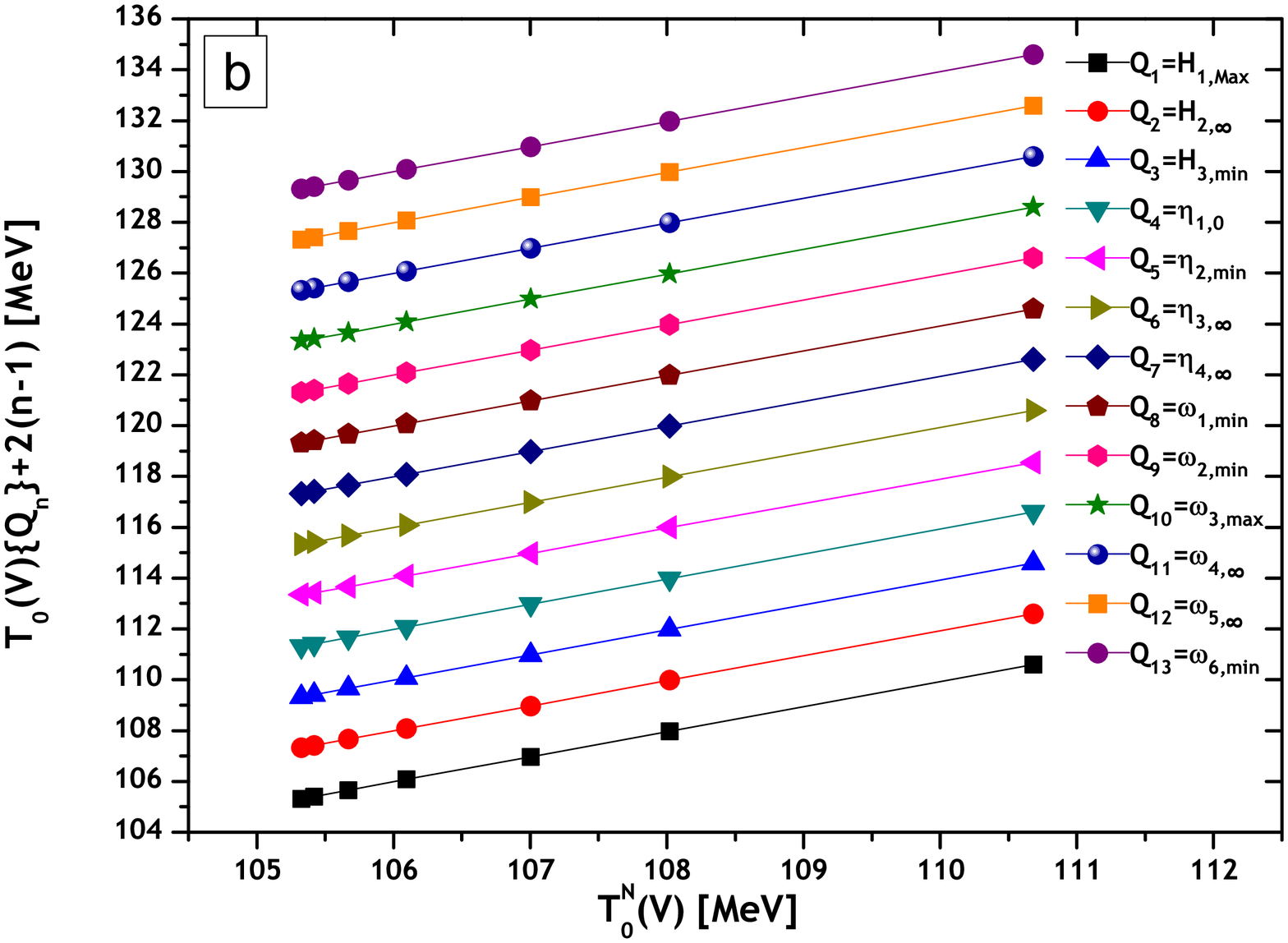}
  \caption{(a) Zoom of the transition region in different Octosis vs Temperature at $V=100fm^{3}$ - (b) Correlation scatter plot between $T_{0}(V)\lbrace {Q}_{n}\rbrace + (n-1)(2MeV) $ and $ T_{0}^{N}(V)$ for different volumes $({Q}_{1}=(\mathcal{H}_{1})_{Max}, {Q}_{2}=(\mathcal{H}_{2})_{\infty}, {Q}_{3}=(\mathcal{H}_{3})_{min}, {Q}_{4}=(\eta _{1}) _{0}, {Q}_{5}=(\eta _{2}) _{min}, {Q}_{6}=(\eta _{3}) _{\infty}, {Q}_{7}=(\eta _{4})_{\infty}, {Q}_{8}=(\omega_{1})_{min},{Q}_{9}=(\omega_{4})_{min},{Q}_{10}=(\omega_{2})_{Max},{Q}_{11}=(\omega_{5})_{\infty},{Q}_{12}=(\omega_{3})_{\infty},{Q}_{13}=(\omega_{6})_{min}.$)}
\end{figure}
\vspace{-2mm}
\section{Results and Discussion}
The plots displayed in Fig1(a),Fig1(b) and Fig2(a) represent the variations of $\mathcal{H}_{1,2,3}(T,V)$,$\eta_{1,2,3,4}(T,V)$,$\omega_{1,2,3,4,5,6}(T,V)$ as a function of temperature at fixed volume: $V=100fm^3$. They show a same global behavior outside the transition region and a different oscillatory behavior in it, however the structure of the particular points are much different. The broadening effect of the transition region with decreasing volume is also observed.
Using the method for locating the finite volume transition point, as developed in \cite{Ladrem2015} , we shall show how this finite volume phase transition, clearly manifests itself as a particular point in each cumulant. Our strategy, we use consists of finding a judicious point where the temperature $T_{0}(V)$, seemingly tends to the bulk $T_{0}(\infty )$\ with increasing volume: $\lim\limits_{(V)\to \infty}T_{0}(V)=T_{0}(\infty )$. Also $T_{0}(V)$ must be highly correlated with the temperature $T_{0}^{N}(V)$ obtained from the natural definition of the finite volume transition point as the extrema of thermal susceptibility $\chi _{T}$, and specific heat $c_{T}$ as given in table (2) of \cite{Ladrem2015}, meaning that the scatter plot should be described by a linear function of the form $ T_{0}(V)={\lambda} T_{0}^{N}(V)+{\nu}$. The first constant $\lambda$ represents the correlation factor and should be close to unity, while the second constant $\nu$ must be close to zero.
The definition of $T_{0}(V)$ is not arbitrary but very difficult analytically and differs according to the quantity being considered. After a careful analysis of the H.O.C plots, we identify the same particular points as defined in \cite{Ladrem2015} . We have analyzed several particular points and only good candidates are considered in this work. If a particular point is considered as a good finite volume transition point, one would expect that its correlation with natural transition point satisfies the same three criteria as in \cite{Ladrem2015} . It can be noted from the scatter plots Fig.2(b) that these points are closely distributed on straight lines, reflecting a strong linear relationship between the two sets of data and the numerical values of the slopes are close to unity as expected (\ref{tab:1}) with a mean value $ <\lambda>=0.98515$. In order to avoid overlapping between fitting curves in Fig.2(b) and to allow a clear representation on the same graph, we have added a shift of 2 MeV between each two consecutive curves.
From this, we deduce that the numerical values of temperature $T_{0}(V)$ of the various transition points, are comparable with an accuracy less than $2\%$, compared to the temperatures $T_{0}^{N}(V)$ extracted using conventional procedures.
The local maximum point in $\mathcal{H}_{1}$ becomes a singularity point in $\mathcal{H}_{2}$ and a local minimum in $\mathcal{H}_{3}$. The zero (inflection) point in $\eta_{1}$ becomes a local minimum point in $\eta_{2}$, a singularity point in $\eta_{3}$ and in $\eta_{4}$. The local minimum point in $\omega_{1}$ becomes a local maximum point in $\omega_{2}$, a singularity point in $\omega_{3}$, a local minimum point in $\omega_{4}$, a singularity point in $\omega_{5}$ and a local minimum in $\omega_{6}$. Moreover, the obvious change in the sign, observed in our results, is in agreement with the results obtained by other models \cite{SignCumulant} .
\begin{table}[h]
\caption{Correlation factor values obtained from linear fitting in Fig2.b}
\label{tab:1}
\tabcolsep6pt\begin{tabular}{lcccccccc}
\hline
\tch{1}{c}{b}{$\mathcal{H}(T,V)$ }  & \tch{1}{c}{b}{T. Point}  & \tch{1}{c}{b}{$\lambda$ }  & \tch{1}{c}{b}{$\eta(T,V)$}  & \tch{1}{c}{b}{T. Point}
& \tch{1}{c}{b}{$\lambda$ } & \tch{1}{c}{b}{$\omega(T,V)$ }  & \tch{1}{c}{b}{T. Point }
& \tch{1}{c}{b}{$\lambda$ }   \\
\hline
$\mathcal{H}_{1}(T,V)$ &  $(\mathcal{H}_{1})_{Max}$ & 0.98456 &$\eta_{1}(T,V)$ &$(\eta _{1}) _{0}$ & 0.98627& $\omega_{1}(T,V)$    &$(\omega_{1})_{min}$  & 0.98456\\
$\mathcal{H}_{2}(T,V)$    &$(\mathcal{H}_{2})_{\infty}  $  & 0.98625  & $\eta_{2}(T,V)$    &   $(\eta _{2}) _{min}$  & 0.98627 & $\omega_{2}(T,V)$    &$(\omega_{2})_{Max}$  & 0.98456\\
$\mathcal{H}_{3}(T,V)$    &  $(\mathcal{H}_{3})_{min}$  & 0.98456  & $\eta_{3}(T,V)$    &$(\eta _{3}) _{\infty}$  & 0.98495 & $\omega_{3}(T,V)$ & $(\omega_{3})_{\infty}$ & 0.98495 \\
 &  &  &$\eta_{4}(T,V)$ &$(\eta _{4})_{\infty}$ & 0.98630 & $\omega_{4}(T,V)$    &$(\omega_{4})_{min}$  & 0.98456 \\
& & &  &  & &     $\omega_{5}(T,V)$    &$(\omega_{5})_{\infty}$  & 0.98461\\
& & &  &  & &     $\omega_{6}(T,V)$ &$(\omega_{6})_{min}$ & 0.98456\\
\hline
\end{tabular}
\end{table}
\vspace{-2mm}
\section{Conclusion}
\vspace{-1mm}
From the new reformulation of the cumulant ratios and using our hpdf and the $\mathscr{L}_{m,n}$-Method, three finite volume H.O.C of the order parameter are calculated and studied. We have noticed that these H.O.C and their ratios, associated to the thermodynamical fluctuations of the order parameter, in QCD behave in a particular enough way revealing pronounced oscillations in the transition region. The sign structure and the oscillatory behavior of these in the vicinity of the deconfinement phase transition point might be a sensitive probe and may allow to elucidate their relation to the QCD phase transition point. In the context of our model, we have shown that the finite volume transition point is always associated to the appearance of a particular point in whole cumulants under consideration. A detailed Finite Size Scaling (FSS) analysis of the results has allowed us to locate the finite volume transition points and extract accurate values of their temperatures $T_{0}(V)$. In addition to natural definition of finite volume transition point as the extrema of thermal susceptibility $\chi _{T}$
 and specific heat $c_{T}$, we have shown that the true finite volume transition point manifests itself as a different particular point according to the quantity considered, namely as,\\
(1) a local maximum point:$(\mathcal{H}_{1})_{Max}$,$(\omega_{2})_{Max}$
\hspace{39mm}
(2) a zero (inflection) point :$(\eta _{1}) _{0}$\\
(3) a local minimum point:$(\mathcal{H}_{3})_{min}$,$(\eta _{2}) _{min}$,$(\omega_{1,4,6})_{min}$
\hspace{28mm}
(4) a singularity point:$(\mathcal{H}_{2})_{\infty}$,$(\eta _{3,4}) _{\infty}$,$(\omega_{3,5})_{\infty}$\\
These results are in complete agreement with those obtained from the FSS analysis of lower order cumulants ratios that published in \cite{Mhamed2014,Ladrem2015} .





\vspace{-2mm}
\begin{thebibliography}{99}
\vspace{-1mm}
\normalsize
\bibitem{Qin2014} G. Y. Qin, \ Nuclear Physics A \textbf{931},165-175,(2014).
\bibitem{MITBM} A. Chodos et al., Phys. Rev. D \textbf{9} (1974) 3471; J. Cleymans  Phys. Rep. \textbf{130} (1986) 217
\bibitem{Ladrem2005} M. Ladrem, A. Ait-El-Djoudi, Eur. Phys. J. C \textbf{44} 257 (2005) (arXiv:0412407v1[hep-ph]).
\bibitem{Herbadji2007} S. Herbadji, Magister thesis in theoretical physics, Ecole Normale Sup\'{e}rieure-Kouba, Algiers,Algeria (2007).
\bibitem{LZH2011} M. Ladrem, Z. Zaki-Al-Full \& S. Herbadji, AIP CP, \textbf{1343}, 492,(2011); Ibid, \textbf{1370}, 226,(2011) .
\bibitem{Mhamed2014} M.A.A. Ahmed, Master thesis in theoretical physics, Taibah University, Al-Madinah Al-Mounawwarah, KSA (2014).
\bibitem{Ladrem2015} M. Ladrem, M.A.A. Ahmed, Z. Al-Full, S. Cherif, Eur. Phys. J. C \textbf{75}, 431, (2015)  (arXiv:1509.00954 [hep-ph]).
\bibitem{Risken1989} Hannes Risken,\textsl{The Fokker-Planck Equation }\ (Springer-Verlag, 1989).
\bibitem{SignCumulant} M.A. Stephanov Phys. Rev. Lett. \textbf{107} (2011) 052301, B. Friman and al. Eur. Phys. J. C \textbf{71} (2011) 1694.


\end{thebibliography}

\end{document}